\documentclass[a4paper, 10pt, twocolumn]{article}

\usepackage{latexsym}
\usepackage{amsmath}
\usepackage{amsfonts}
\usepackage{amssymb}
\usepackage{graphicx}
\usepackage{color}

\begin{document}
\title{Co-evolving  agents subject to  local versus nonlocal barycentric interactions}

\author{{\,}\\{\large Max-Olivier Hongler} \\ {\scriptsize {\texttt Ecole Polytechnique F\'{e}d\'{e}rale de Lausanne, CH-1015 Lausanne, Switzerland}}\\ $\,$\\ Roger Filliger\\ {\scriptsize {\texttt Bern University of Applied Sciences, CH-2501 Biel, Switzerland}}\\ $\,$\\ Olivier Gallay\\ {\scriptsize {\texttt IBM Zurich Research Laboratory, CH-8803 Rueschlikon, Switzerland}}\\{\,}\\}
\maketitle
\abstract{\noindent The mean-field dynamics of a collection of stochastic agents with local versus nonlocal interactions is studied via analytically soluble models. The nonlocal interactions result from a barycentric modulation of the observation range of the agents.  Our modeling framework is based on a discrete two-velocity Boltzmann dynamics which can be analytically discussed.
Depending on the span and the modulation of the interaction range, we analytically observe a transition from a purely diffusive regime without definite pattern to a flocking evolution represented by a solitary wave traveling with constant velocity.}

\vspace{0.5cm}

\noindent {\bf Keywords}: Self-organized systems, Transport processes, Kinetic theory.

\maketitle

\section{Introduction}
\noindent  Since several decades, the fascination for flocking dynamics and specifically, the detailed mechanisms which induce a collection of interacting  stochastic agents to exhibit an emergent collective behavior, stimulates a fruitful activity on both experimental and modeling sides. The recent pioneering and truly engrossing contributions from  T. Vicsek et al.~\cite{Vicsek} and F. Cucker and S. Smale~\cite{Smale} unveil, from a simple and synthetic modeling point of view, some of the  basic  features underlying the formation of robust collective motions. 
As intuitively expected, a central role is played by the interactions connecting the agents.
Both the interaction range and the interaction strength are  control parameters which, for critical values, may trigger a flocking (phase) transition. By a flocking pattern,  we understand  the self-organized capability of agents to create, via their mutual  interactions,  a persistent probability density taking the form of a traveling wave. Heuristically, with strong enough interactions, the noise-induced detuning tendency, can be overcome and ultimately produce an emergent synchronized evolution. In  the models  introduced by {Vicsek et al.} and by {Cucker and Smale}, it is indeed the interaction range that directly  tunes the self-organization capability. Alternatively, recent observations for birds demonstrate that a fixed critical number of neighbors rather  than a pure metric spatial range is used by  birds in order to flock together to platoons~\cite{Cavagna, Parisi}. This shows that the interaction range itself may  depend on the instantaneous distribution of agents. Here, we contribute to this widely open topic by proposing a simple class of models, where the interaction range depends in an effective way on the agent distribution.

\noindent Societies of agents can be composed either of  dynamically homogeneous or heterogeneous individuals and each case requires a drastically different modeling approach. Recent analytical, while impressive, results have been derived  for  heterogeneous agents  like  ranked-based  interacting Brownian motions~\cite{Karatzas, Chatterjee, Pal} or bucket brigades dynamics ~\cite{Armbruster, Bartholdi}.  We here focus on homogeneous populations of agents which offer a wide potential for \emph{analytical} approaches. In an homogeneous population,  any randomly selected individual is likely to be a dynamical representative of any other fellow of the society.  This very  basic  feature together with the fact that for large populations, the relative importance of fluctuations diminish, enable, in the thermodynamic limit $N \rightarrow \infty$, to adopt a purely macroscopic description.  This assumes that the normalized distribution of agents can be ``hydrodynamically'' described by a probability density which solves  a deterministic, yet nonlinear, evolution equation. We call the resulting equation the {\it mean-field} (MF) {\it equation}.

\noindent The class of MF models introduced in this contribution shows flocking transitions for large enough interaction ranges and belongs to the rare instances of exactly solvable interacting agent dynamics. The price to be paid for such an analytical objective is to conveniently reduce the individual agents' {\it state space} and {\it decision space}. As a minimal model, we shall consider {\it random two velocity models} involving agents traveling on the real line with two discrete velocities, say $\left\{ v_{\pm} \right\}$ with $v_-<v_+$.  The effective agent autonomous  decisions  will consist in selecting, at non-homogeneous Poisson random times,  one out of the two possible velocities. The velocity updating process will, via density-dependent parameters of the underlying Poisson switching times,  depend on the observation range. This naturally leads to nonlinear Boltzmann-like dynamics, which is commonly adopted  for traffic  flows modeling~\cite{Bellomo, Helbing, Filliger}.
The observation range, which will be the key  control parameter for our discussion, may depend on the distribution of the whole agent society via a {\it barycentric} modulation. Such a barycentric modulation offers, in an effective manner, the possibility to model agent dynamics with configuration-dependent interaction ranges. This approach  enables us to construct several solvable nonlinear models for which critical interaction modulations -- required for flocking -- can be obtained explicitly.

\noindent  Although different from our present contribution, related MF models exist in the literature. Exactly solvable MF models have been proposed in different contexts including portfolio theory~\cite{Fernholz}, coupled phase oscillators~\cite{Bonilla}, traffic dynamics~\cite{Bellomo, Helbing}, self-propelled organisms~\cite{Toner}, etc. Related MF models with barycentric self-interactions are studied \cite{Balazs} and \cite{Comets}.
 Moreover, the class of models discussed in the present paper is amenable to  a discrete velocity Boltzmann equation of the Ruijgrok-Wu type \cite{RW} and can be hence analytically discussed.
   As originally discussed in \cite{Gutkin, McKean}, for large populations of agents, our discrete velocities Boltzmann dynamics can be made to converge towards the generalized  Burgers equation for which exact results \emph{and} stability issues for traveling waves can be discussed explicitly,~\cite{Nishihara}.

\section{Two velocity agent model}
Consider a collection ${\cal A}$ composed  of $N$ autonomous agents  which are  in a migration process on the real line $\mathbb{R}$. At any time $t\in \mathbb{R}^{+}$, we assume that the complete population  is composed of two types of agents ${\cal A}_+$ and ${\cal A}_-$, (${\cal A}={\cal A}_+\cup {\cal A}_-$),  characterized  by two associated migration velocities $v_+$ and $v_-$ on $\mathbb{R}$.  Agents traveling  with velocity $v_+$ (resp. $v_-$) belong to ${\cal A}_+$ (resp. ${\cal A}_-$). The time-dependent positions of the agents $X_k(t)$, $k=1,2,...,N$, can be written as a set of coupled stochastic differential equations (SDEs):
\begin{equation}
\label{eq.1}
\dot{X}_k(t) = I_k(\textbf{X}(t)), \qquad k=1,2,...,N,
\end{equation}
\noindent with $\textbf{X}(t)=(X_1(t),...,X_N(t))$ and where $I_k(\textbf{X}(t))$ stands for a two-states Markov process with  state space  $\Omega:= \left\{ v_+, v_-\right\}$.  The associated  transition rates of agent $k$, $\alpha=\alpha\big(X_k(t),\textbf{X}(t)\big)>0$ from $v_+$ to $v_-$ and $\beta=\beta\big(X_k(t),\textbf{X}(t)\big)>0$ from $v_-$ to $v_+$ are here state-dependent. For a given configuration $\textbf{X}(t)$ at time $t$, they represent the inverse of the average sojourn times of agent $k$ in the velocity states.
This dependency of the switching rates on the agent population effectively allows agent $k$ to observe his/her environment and to react accordingly. In the sequel, we consider large populations of homogeneous agents, {\it i.e.} $N\rightarrow \infty$, so that a MF description of the dynamics holds. Using a phenomenological MF representation, we shall write $P_{+}(x,t\mid x_0)$ (resp. $P_{-}(x,t\mid x_0)$) for the conditional probability density to find agents at position $x$ at time $t$ with velocity $v_{+}$ (resp. $v_{-}$), knowing that at time $t=0$ the density was given by the initial distribution $p_+(x)$ (resp. $p_-(x)$). Due to the assumed  Markov character of the transitions, the nonlinear evolution for $P_{+}(x,t\mid x_0)$ and $P_{-}(x,t\mid x_0)$ can be written as a discrete Bolzmann-type equation ({\it n.b.} we drop the $x_0$-dependence for typographic ease):
$$
\dot{P}_{\pm}(x,t) + v_{\pm}  \partial_{x}P_{\pm}(x,t) =  \quad \quad \quad \quad \quad \quad \quad \quad \quad \quad
$$
\begin{equation}
\label{eq.3}
 \quad \quad \quad\mp \alpha(x,t) P_{+}(x,t)\pm \beta(x,t) P_{-}(x,t),
\end{equation}
\noindent with interaction kernels:
\begin{eqnarray}
\label{eq.4}
\,\alpha(x,t) =
 \alpha - \int_{x-U}^{x+V} g\left[z- \langle X(t) \rangle\right] P_{-}(z,t) dz ,\\
\label{eq.5}
\beta(x,t) =  \beta+ \int_{x-U}^{x+V} g\left[z- \langle X(t) \rangle\right] P_{+}(z,t) dz,
 \end{eqnarray}
\noindent and where now $\alpha$ and $\beta$ are two positive constants,  $g(x)\geq 0$  is a smooth function and integrable with respect to the $P_+$ and $P_-$ probability densities and where
 \begin{equation}
\label{eq.6}
\langle X(t) \rangle  = \int_{\mathbb{R} }x \left[ P_{+}(x,t) + P_{-}(x,t) \right] dx
\end{equation}
is the {\it barycenter}. According to eqs.~(\ref{eq.3}-\ref{eq.6}),  agents modify their velocity by the following dynamic rules:
\noindent
\begin{itemize}
  \item[{\it (i)}]  {\it Agents with velocity $v_+$}. An agent at position $x$ and with velocity $v_+$  changes spontaneously  to $v_-$ with constant rate $\alpha$. This agent is allowed to observe the environment in the observation interval ${\cal O}= [x-U, x+V]$. The $\alpha$-rate is reduced in a weighted proportion to the number of $v_-$ agents present in ${\cal O}$. When $g \neq {\rm constant}$, the weight is modulated by the barycentric position $\langle X(t) \rangle$ of the population via the $g$-dependence.
  \item[{\it (ii)}]  {\it Agents with velocity $v_-$}.  An agent at position $x$ and with velocity $v_-$  changes spontaneously  to $v_+$ with constant rate $\beta$. This agent is allowed to observe the environment in the observation interval ${\cal O}= [x-U, x+V]$. The $\beta$-rate is enlarged  in a weighted proportion to the number of $v_+$ agents present in ${\cal O}$. When $g \neq {\rm constant}$, the weight is modulated by the barycentric position $\langle X(t) \rangle$ of the population via the $g$-dependence.

\end{itemize}

\noindent The following remarks can be pointed out:

 \begin{enumerate}

 \item [{\it (1)}] Agents interact non-locally with their neighbors in two ways: {\it (a)} via the finite extension of the ${\cal O}$ interval and {\it (b)} via the $g$-modulation which  depends on the barycenter $\langle X(t) \rangle$. Hence, for ${\cal O} \neq \mathbb{R}$, one could find an inconstancy between the two mechanisms. This however is not so as the agents may simultaneously base their decisions on  real-time information delivered by  physically different sensors like {\it vision, sound, olfaction}, etc.

 \item [{\it (2)}] Observe that the sum of the right-hand-sides of the two equations given in eqs.~(\ref{eq.3}) is equal to zero. This is an expression of the continuity equation which guarantees  the conservation of the number of agents.

\item [{\it (3)}] Using the rescaling $v_{\pm} \mapsto v_{\pm} /\epsilon$, $\alpha \mapsto \alpha/\epsilon^{2}$ and $\beta \mapsto \beta /\epsilon^{2}$ in eqs.~(\ref{eq.3}), our two velocity model converges for $\epsilon \rightarrow 0$ to a Burgers' type dynamics of the form:
$$
\partial_t \Psi(x,t) = \quad\quad\quad\quad\quad\quad\quad\quad\quad\quad\quad\quad\quad\quad\quad\quad\quad\quad 
$$
$$
-\partial_x \left\{ \left( f(x,t) + \gamma \left[ \int_{x-V}^{x+U} \Psi(z,t ) \, dz \right] \right)\Psi(x,t)   \right\}
$$
\begin{equation}
\label{FP}
+ {D \over 2}\partial_{xx}^{2} \Psi(x,t),
\end{equation}
\noindent which is a MF nonlinear Fokker-Planck equation for  interacting diffusion processes driven by White Gaussian Noise, ~\cite{Gambetta-Perthame, Hongler-Streit}. We exemplify in appendix A how to transfer our findings valid for the Boltzmann-type dynamics  to this Burgers' type limiting model.

\end{enumerate}

\noindent We now explicitly give the solutions to the set of two non-linearly coupled equations given in eqs.~(\ref{eq.3}) for the following regimes:
\begin{itemize}
  \item[\textit{(A)}]  \emph{Follow the immediate leaders} : $U=0$, $V= \varepsilon$, with $\varepsilon$ infinitesimal small, and $g\equiv 1$,
  \item[\textit{(B)}]  \emph{Follow all the leaders}:  $U=0$, $V= \infty$ and $g\equiv 1$,
  \item[\textit{(C)}]  \emph{Follow the leaders with barycentric modulation}:  $U=0$, $V= \infty$ and $g= g(x- \langle X(t) \rangle)$.
\end{itemize}
\noindent Studying the regimes $(A)$ and $(B)$, we quantify the importance played by the size of the observation interval ${\cal O}$. Comparing the regimes $(B)$ and $(C)$ we may appreciate how the influence of the $g$-barycentric modulation will give rise to flocking phase transitions.

\section{Discrete velocity dynamics}
\label{RWSEC}

\noindent  Using a simple rescaling of the state and time variables ({\it e.g.} see ~\cite{Filliger} for details), we can without loss of generality set $v_+=-v_-=1$ (and similarly, $\gamma=1$ in eq.~(\ref{FP})) and rewrite eqs.~(\ref{eq.3}) in canonical form:
$$
\dot{P}_{\pm}(x,t) \pm \partial_{x}P_{\pm}(x,t) =  \quad\quad\quad\quad\quad\quad\quad\quad\quad\quad\quad$$
\begin{equation}
\label{eq.8}
\quad\quad\quad\quad\mp \alpha(x,t) P_{+}(x,t)\pm \beta(x,t) P_{-}(x,t).
\end{equation}
Based on eq.~(\ref{eq.8}), we now explore the above mentioned regimes.
\subsection*{\emph{(A)} Follow the immediate leaders}
\label{SRW}

\noindent In the myopic case where $V=\varepsilon$ is very small and $U=0$, we may Taylor expand up to first order the quadrature in eqs.~(\ref{eq.4}) and (\ref{eq.5}). The resulting dynamics from eqs.~(\ref{eq.8}) reads as:
$$
\dot{P}_{\pm}(x,t) \pm \partial_{x}P_{\pm}(x,t) = \quad\quad\quad\quad\quad\quad\quad\quad\quad\quad\quad\quad\quad\quad
$$
\begin{equation}
\label{eq.9}
\quad\quad{\pm}2\varepsilon P_{+}(x,t) P_{-}(x,t)  \mp \alpha P_{+}(x,t) \pm \beta P_{-}(x,t).
\end{equation}
\noindent This is an exactly solvable, discrete Boltzmann-type equation discovered  by Th. Ruijgrok and T. T. Wu, ~\cite{RW}. Potential applications for this type of dynamics have been recently considered  in~\cite{Hashemi, Gallay}. Using the boundary conditions $\lim_{|x|\rightarrow \infty} P_{\pm}(x,t) =0$, the solution reads as:
 \begin{eqnarray}
 \label{eq.12}
P_{+}(x,t) = -\beta  + \partial_t \log H(x,t)   - \partial_x \log H((x,t),\\
\label{eq.12b}
 P_{-}(x,t) = \phantom{-}\alpha - \partial_t \log H(x,t)   - \partial_x \log H(x,t),\,\,\,
\end{eqnarray}
\noindent where $H(x,t) $ solves the linear {\it Telegraphist equation}:
\begin{equation}
\label{eq.14}
\partial_{tt} H(x,t)   - \partial_{xx} H(x,t) - \alpha \beta  H(x,t) =0
\end{equation}
and whose explicit solution is recalled in the appendix B.

\noindent The diffusive solution does not converge towards a finite stationary density, implying that,  for any initial condition, the agents will ultimately be spread over the whole line $\mathbb{R}$. This explicitly indicates that local interactions are \emph{not} strong enough to create a cooperative motion ({\it i.e.} no flock is formed).
\subsection*{\emph{(B)} Follow all the leaders}

\noindent Instead of the myopic regime $(A)$ given in eqs.~(\ref{eq.9}), let us now consider  case $(B)$, namely:
$$
\partial_t P_{\pm}(x,t) \pm \partial_{x}P_{\pm}(x,t) = \mp \left[ \alpha - \int_x^{\infty} P_{-} dz \right] P_{+}(x,t)\quad\quad
$$
\begin{equation}
\label{eq.15}
\quad\quad\quad\quad\quad\quad\quad\quad\,\,\pm \left[ \beta +  \int_x^{\infty} P_{+}dz \right] P_{-}(x,t).
 \end{equation}
\noindent Introducing the notations
\begin{equation}
\label{FPRGPR}
F_{\pm}(x,t) = \int_x^{\infty} P_{\pm}(z,t)\, dz,
\end{equation}
\noindent  we can rewrite eqs.~(\ref{eq.15}) as:
$$
\partial_{xt} F_{\pm}(x,t) \pm \partial_{xx}F_{\pm}(x,t) =\pm \partial_x \left[F_{-}(x,t) F_{+}(x,t)  \right] \quad\quad\quad
$$
\begin{equation}
\label{eq.20}
\quad\quad\quad\quad\quad\quad\quad\quad\quad\mp \alpha \partial_xF_{+}(x,t)  \pm \beta \partial_x F_{-}(x,t).
 \end{equation}
 \noindent  After integration with respect to $x$, eqs.~(\ref{eq.20}) exhibit the same structure than eqs.~(\ref{eq.9}) for the fields $F_{+}(x,t)$ and $F_{-}(x,t)$. Here however, the boundary conditions are:
 \begin{eqnarray}
\label{eq.22}
 \lim_{x\rightarrow \infty} F_{\pm}(x,t) =0, \quad\,\, {\rm and} \quad\,\,    \lim_{x\rightarrow -\infty} F_{\pm}(x,t)= \rho_{\pm}
\end{eqnarray}
\noindent and the normalization imposes that $(\rho_{+}+\rho_{-})=1$. With these boundary conditions, the solution reads as,~\cite{RW}:
\begin{equation}
\label{FFF}
F_{\pm}(x,t) = {\rho_{\pm} \over 2 } \left\{1 - \tanh\left[\frac{ x -wt}{4}\right]\right\}, 
\end{equation}
\noindent with
\begin{equation}
\label{GGG}
2\rho_{\pm}= 1\pm w>0,
\end{equation}
\noindent and where $w$ is  the velocity defined by :
\begin{equation}
\label{CELOS}
w=  \sqrt{ \left[\alpha+ \beta-1 \right]^{2} + 4 \beta}  - (\alpha + \beta).
\end{equation}
\noindent From eq.~(\ref{FFF}), the probability densities in eqs.~(\ref{eq.15}) exhibit the form  of solitary waves:
\begin{equation}
\label{DENSITY}
P_{\pm}(x,t) = -\partial_xF_{\pm}(x,t) ={\rho_{\pm} \over  8\, {\cosh^{2} \left(\frac{x-wt}{4}\right) }}\cdot
\end{equation}
\noindent The solutions in eqs.~(\ref{DENSITY}) explicitly show that in case $(B)$,   the long-range imitation process generates stationary, finite probability densities. Hence contrary to what happens in the myopic regime $(A)$, here {\it flocking} results from the long-range of the agents' mutual interactions.
%

\subsection*{\emph{(C)} Follow the leaders with barycentric modulation}

\noindent The above analysis of the regimes $(A)$ and $(B)$ suggest that for a certain critical interaction strength, one should be able to observe a phase transition from a purely diffusive regime with no stationary patterns to a stationary dispersive flocking regime. A very simple, yet exact solution to this problematic is shown here, where we now introduce an explicit barycentric modulation function. Recent contributions using barycentric interactions are discussed in \cite{Balazs, Comets, Frank, Pal}. The class of models we consider here are of the following form:
$$
\dot{P}_{\pm}(x,t) {\pm}  \partial_{x}P_{\pm}(x,t) =  \mp P_{+}(x,t) \left[ \alpha - \mathbb{F}_-(x,t) \right] \quad\quad\quad\quad
$$
\begin{equation}
\label{eq.27}
\quad\quad\quad\quad\quad\quad\quad\quad\quad\quad\quad\,\,{\pm} P_{-}(x,t) \left[ \beta + \mathbb{F}_+(x,t) \right],
\end{equation}
\noindent with
 \begin{eqnarray}
\label{eq.28}
\mathbb{F}_{\pm}(x,t)= \int_{x}^{\infty} g\left[z- \langle X(t) \rangle\right] P_{\pm}(z,t) dz,
\end{eqnarray}
\noindent and where $\langle X(t) \rangle$, defined in eq.~(\ref{eq.6}), is the barycenter position at time $t$.

\noindent Let us from now on consider traveling wave solutions to eqs.~(\ref{eq.27}) of the form $f(x-wt)$ for some velocity $w$. In such a traveling wave regime and for large $t$,  we have $\langle X(t) \rangle  = w\, t$. Introducing the variable $\xi = (x-w\,t)$ we may, by a slight abuse of notations, rewrite $P_{\pm}$ in this regime as $P_{\pm}(x,t)=P_{\pm}(\xi)$ and eq. (\ref{eq.6}) now reads as:
 \begin{equation}
\label{eq.29b}
0  = \int_{\mathbb{R} }\xi \left[ P_{+}(\xi) + P_{-}(\xi) \right] d\xi,
\end{equation}
meaning that in the traveling wave case, the stationary distribution, as seen from the center of mass,
has zero mean.
 As the right-hand-sides in eqs.~(\ref{eq.27}) sum to zero, we immediately find, by simple integration with respect to $\xi$, the following relation:
\begin{equation}
\label{EQUI}
(1-w)P_{+}(\xi) = (1+w) P_{-}(\xi) + \kappa,
\end{equation}
\noindent where the celerity of the center of mass $w$ will be chosen below, so as to be consistent with the agent distribution. The integration constant $\kappa$ will be set to zero in order to match natural ({\it i.e.} vanishing) boundary conditions at infinity. Therefore, we shall have:
\begin{equation}
\label{INTERBAR}
P_{-}(\xi) = \frac{1-w}{1+w} P_{+}(\xi).
\end{equation}
Using eq.(\ref{INTERBAR}) and writing eqs.~(\ref{eq.27}) with the variable $\xi$, we obtain the MF evolution equation:

$$
\partial_{\xi} P_{+}(\xi) = \quad\quad\quad\quad\quad\quad\quad\quad\quad\quad\quad\quad\quad\quad\quad\quad\quad\quad\quad\quad
$$
\begin{equation}
\label{SOL/BAR}
P_{+}(\xi) \Big(\frac{2}{ 1+w} \int_{\xi}^{\infty} g(\zeta)P_{+}(\zeta) d\zeta  +\frac{\beta}{1+w}-
\frac{\alpha}{1-w}\Big).
\end{equation}

\noindent In view of our analytical objective we now introduce the class of symmetric interaction modulations:
\begin{equation}
\label{MOD/BAR}
g(x) = \Delta \cosh^{\eta}(x),
\end{equation}
\noindent with $\Delta >0$ and $\eta \in \mathbb{R}$, together with the {\it Ansatz}:

\begin{equation}
\label{eq.35}
P_{+}(\xi) = {\cal N}(m) \cosh^{m}(\xi),
\end{equation}
\noindent where $m<0$ is some negative constant and where ${\cal N}(m)$ stands for the probability normalization, explicitly :
\begin{equation}
\label{CALN}
{\cal N}(m) = {\sqrt{\pi}  \, \Gamma(|m|/2) \over 2^{|m|} \Gamma\left[(|m|+1)/2\right]}\cdot
\end{equation}

\noindent Introducing eqs.~(\ref{MOD/BAR}) and (\ref{eq.35}) into  the integral equation~(\ref{SOL/BAR}), an elementary calculation shows that eq.~(\ref{eq.35}) is actually a solution provided we impose:
\begin{equation}
\label{IMPOBAR}
m + \eta  = -2,
\end{equation}
\noindent in which case we end up with the relation:
\begin{equation}
\label{FINAL/BAR}
\left[ m+ {2 \Delta {\cal N}(m) \over1+w}  \right]\tanh(\xi) =  \frac{2 \Delta {\cal N}(m)+ \beta}{1+w}- \frac{\alpha}{1-w}\cdot
\end{equation}
\noindent Clearly, eq.~(\ref{FINAL/BAR}) is realized only if the following equalities hold
\begin{equation}
\label{SOL/BAR/FIN}
{m \over {\cal N}(m)}= -{2\Delta\over 1+w} \quad {\rm and}  \quad {1-w\over 1+w} = {\alpha \over\beta + 2 \Delta {\cal N}(m)},
\end{equation}
\noindent implying the self-consistency equation for the traveling speed:
\begin{equation}
\label{CELEROS}
w = 1+ {2 \alpha \Delta {\cal N}(m) \over m\left[\beta + 2 \Delta {\cal N}(m) \right] }\cdot
\end{equation}
\noindent Hence, given the positive input modeling  data  $\alpha$, $\beta$ and $\Delta$, together with the modulation $\eta$ of the barycentric interaction $g$, eq.~(\ref{IMPOBAR}) determines  the probability density  decay   $m$ and eq.~(\ref{CELEROS}) determines the resulting solitary wave velocity $w$. Note that the normalization factor in eqs.~(\ref{eq.35}, \ref{CALN}) exists only for $m<0$, or equivalently for $\eta > -2$, which yields the critical power $\eta_c$ for the range of the modulation decay:
\begin{equation}
\label{IMPOBAR2}
\eta > \eta_c=-2.
\end{equation}

\noindent The condition on the  modulation range power  $\eta$ in eq.~(\ref{IMPOBAR2}) leads to conclude  that for  strongly  localized  interactions, $\eta < \eta_c$, no finite stationary probability density exists, and this indicates that no stable flocking does emerge from the agent interactions.  Conversely, when  $\eta > \eta_c$, a finite stationary probability density exists, thus showing that comparatively long-range interactions ultimately  drive the population to a persistent flocking behavior.

\section{Conclusion and perspectives}

\noindent The models we have exposed in this contribution simultaneously involve two types of nonlinear sources: first an imitation interaction mechanism of the quadratic type which reflects the observation of the state of the agents by their fellows and secondly a barycentric modulation for the strength of the interactions. The possibility to derive an exactly solvable model for the flocking transition dynamics relies partially in the fact that one nonlinearity can be removed. Indeed, the quadratic nonlinearity can be ``removed'' by using a logarithmic transformation, but the second type of nonlinearity remains as it is truly intrinsic.  In actual agent models like birds or fishes, interactions are characterized by both of these nonlinearities: by imitation (stylized here by the quadratic nonlinearity) and by tuning the observation range according to the agents' local density (effectively stylized here by a barycentric modulation). Another important simplification allowing the derivation of  exactly solvable models is the mean-field limit adopted here. In this limit, the law of large numbers reduces the influence of the fluctuations to insignificancy and this leads to deterministic evolution equations for the agent distribution. For many applications in perspective, finite-size population effects will manifest via the presence of fluctuations ({\it i.e.} a {\it mesoscopic description}). Noise will definitely affect the dynamics and could potentially destroy the flocking capability of the agents. We hope that the basic models discussed in this paper could provide some analytical clues for this truly challenging issue.
\noindent

\section*{Appendix A}

\noindent The complete program performed in this paper can be repeated for the Burgers' dynamics given in eq.(\ref{FP}). Let us only focus here on the interactions given in case $(C)$ for which the dynamics reads as:
$$
\partial_t \Psi(x,t) = -\partial_x \left[ - \Psi(x,t)  \int_{x}^{\infty}g\left(z- \langle X(t) \rangle  \right)\Psi(z,t)  dz \, \right. 
$$
\begin{equation}
\label{GENBU}
\left. - {D \over 2} \partial_x \Psi(x,t)  \right],
\end{equation}
\noindent where:
\begin{equation}
\label{AVOSS}
\langle X(t) \rangle= \int_{\mathbb{R}} x \Psi(x,t) dx.
\end{equation}

\noindent Concerned with traveling wave solutions with constant celerity $w$, we introduce the notation $\xi =( x- wt)$. Accordingly, eq.~(\ref{GENBU}) takes the form:

\begin{equation}
\label{BUSTA}
0 = \partial_{\xi}\left[ \Psi(\xi)\left\{ w-  \int_{\xi}^{\infty} g\left(z  \right)\Psi(z)  dz \right\}  - {D \over 2} \partial_{\xi} \Psi(\xi)
\right]
\end{equation}
\noindent and $w$ is implicitly determined by the equation:
\begin{equation}
\label{Condition}
\int_{\mathbb{R}} \xi \Psi(\xi) d\xi =0.
\end{equation}
\noindent Eq.(\ref{BUSTA}) can be reduced to the form:
\begin{equation}
\label{COMPAT}
{D \over 2} \, \partial_{\xi} \log \left[ \Psi(\xi) \right] = \left\{ w-  \int_{\xi}^{\infty} g\left(z \right)\Psi(z)  dz \right\} + \kappa,
\end{equation}
\noindent where $\kappa$ is a constant which will be taken to be zero as no secular probability flow exists in the stationary regime. Assume now a symmetric barycentric modulation of the form:
\begin{equation}
\label{MODULO}
g(x) = \Delta \cosh^{\eta}(x),
\end{equation}
\noindent with $\Delta >0$ a constant and $\eta \in \mathbb{R}$.  In view of eq.(\ref{MODULO}), we introduce the {\it Ansatz}:
\begin{equation}
\label{ANSATZ}
\Psi(\xi) = {\cal N}(m) \cosh^{m}(\xi),
\end{equation}
\noindent where again, for $m<0$, the normalization  factor   ${\cal N}(m)$ is given by eq.~(\ref{CALN}). Note from eq.~(\ref{ANSATZ}) that $\Psi(\xi) = \Psi(-\xi)$ and therefore eq.~(\ref{Condition}) is automatically satisfied. Using eqs.~(\ref{ANSATZ}) and (\ref{CALN}) into eq.~(\ref{COMPAT}), we end up with:

\begin{equation}
\label{CALCUL}
{D \over 2}\,  m\tanh(\xi) =  \left[w -      \Delta {\cal N}(m)\int_{\xi}^{\infty} \cosh^{\eta + m}(\xi)  d\xi \right].
\end{equation}
\noindent   Using the identity  $\int \cosh(x)^{-2} dx = 1-\tanh(x)$,  we see   that eq.(\ref{CALCUL}) is solved provided we have:
\begin{equation}
\label{SOLBU}
{m \over {\cal N}(m)}= {2\Delta \over D} \quad {\rm and} \quad w =\Delta {\cal N}(m). \end{equation}
\noindent From eqs.~(\ref{SOLBU}), one concludes that  for  $\eta >-2$, a traveling solitary wave with celerity $w$ is created via  the agents' interactions.  Hence for $\eta > -2$ the  {\rm flocking} mechanism is triggered.  Conversely, for $\eta <-2$, normalization cannot be achieved and this shows that  too  weak  interactions at  long-range preclude the formation of a flock.

\section*{Appendix B}
\noindent The solution $H(x,t) $ to the linear {\it Telegraphist equation},
\begin{equation}
\label{eq.14}
\partial_{tt} H(x,t)   - \partial_{xx} H(x,t) - \alpha \beta  H(x,t) =0
\end{equation}
is of the form,~\cite{Hemmer, RW}:
$$
H(x,t) = {1\over 2}\left[ A(x+t) + A(x-t)\right]\quad\quad\quad\quad\quad\quad\quad\quad
$$
\begin{equation}
\label{BESSEL}
\quad\quad\quad\quad\quad\quad+ {1\over 2} {\cal B}_1(x,t) + {\nu t\over 2} \,  {\cal B}_2(x,t),
\end{equation}
\noindent  where $\nu = \frac{\sqrt{\alpha \beta}}{2}$ and where we have the following definitions:
\begin{equation}
\label{B_1}
{\cal B}_1(x,t)= \int_{x-t}^{x+t} \mathbb{I}_0\left(\nu \sqrt{t^{2} - (x-z)^{2}} \right) B(z)dz,
\end{equation}
\noindent
and
$$
{\cal B}_2(x,t) = \int_{x-t}^{x+t}  \left({ 1 \over \sqrt{t^{2} - (x-z)^{2}}}\right) \quad\quad\quad\quad\quad\quad\quad\quad\quad
$$
\begin{equation}
\label{B_2}
\quad\quad\quad\quad\quad\quad\quad\quad\times\,\, \mathbb{I}_1\left(\nu \sqrt{t^{2} - (x-z)^{2}} \right) A(z)dz,
\end{equation}
\noindent with $\mathbb{I}_n(\cdot)$ being integer-order modified Bessel functions of the first kind and where $B(\cdot)$ and $A(\cdot)$ are short for:
\begin{equation}
\label{INIT1}
B(x) =  {1\over2} \left[P_0(x) - Q_0(x) + \alpha + \beta  \right] A(x),
\end{equation}
\noindent and
\begin{equation}
\label{INIT2}
A(x) =  \exp\left\{ - {1\over2}\int_{0}^{x} \left[P_0(z) + Q_0(z) - \alpha + \beta \right]dz\right\}.
\end{equation}
\section*{Acknowledgements}
This work is in part supported by the Swiss National Foundation for Scientific Research.

\end{document}